\documentclass[aps,prd,fleqn,superscriptaddress]{revtex4}
\usepackage{graphicx,color,natbib}
\usepackage{amsmath,amssymb,amsfonts}

\usepackage{epsfig,epsf}
\usepackage{dcolumn}
\usepackage{float}
\usepackage{bm}

\newcommand{\bse}{\begin{subequations}}
\newcommand{\ese}{\end{subequations}}
\newcommand{\be}{\begin{equation}}
\newcommand{\ee}{\end{equation}}
\newcommand{\bea}{\begin{eqnarray}}
\newcommand{\eea}{\end{eqnarray}}
\newcommand{\ba}{\begin{array}}
\newcommand{\ea}{\end{array}}

\input amssym.def
\input amssym.tex

\usepackage[colorlinks=true, linkcolor=blue, bookmarks=true]{hyperref}

\begin{document}

%
\title{A note on holographic subregion complexity and QCD phase transition}
\author{Mahsa Lezgi\footnote{$\rm{s}_{-}$lezgi@sbu.ac.ir}}
\affiliation{Department of Physics, Shahid Beheshti University G.C., Evin, Tehran 19839, Iran}
\author{Mohammad Ali-Akbari\footnote{$\rm{m}_{-}$aliakbari@sbu.ac.ir}}
\affiliation{Department of Physics, Shahid Beheshti University G.C., Evin, Tehran 19839, Iran}
\begin{abstract}
Using holographic subregion complexity, we study the confinement-deconfinement phase transition of quantum chromodynamics. In the model we consider here, we observe a connection between the potential energy of probe meson and the behavior of its complexity. Moreover, near the critical point, at which the phase transition takes place, our numerical calculations indicate that we need less information to specify a meson in the non-conformal vacuum than in the conformal one, despite the fact that the non-conformal vacuum has larger energy!
\end{abstract}
\maketitle
%
{\textit{\textbf{Introduction}}}: Gauge-gravity duality, or more generally the holographic idea, provides a new framework to investigate various properties of non-perturbative theories within the last two decades. This duality is in fact a strong-week duality and maps a strongly-coupled quantum gauge field theory to a weakly-coupled classical gravity in a higher dimension. This idea has been frequently applied to describe various phenomena in strongly-coupled field theories, for which the standard perturbation method is not applicable, ranging from condensed matter physics to low-energy quantum chromodynamics (QCD), the theory of the strong interactions \cite{CasalderreySolana:2011us,erdmenger}. Central aspects of low energy QCD such as confined phase, confinement-deconfinement phase transition and chiral symmetry breaking have been discussed within the concept of gauge-gravity duality, for instance see \cite{erdmenger} and references therein.

One of the interesting areas of theoretical physics is quantum information theory and, according to holographic idea, an outstanding connection have been developed between quantum information and gravity started by Hubney-Ryu-Takayanagi proposal for entanglement entropy, defined as a measure of quantum correlation of a pure quantum state. Entanglement entropy is one of the important quantities in the context of information theory and fortunately there exists a simple geometrical prescription to describe it and its properties \cite{Takayanagi,Nishioka:2009un} and it passes many tests successfully. 

Another main concept in information theory is quantum complexity. It is defined as the minimum number of unitary operators needed to prepare a target state from a reference state or in other words the difficulty in converting one state to another one \cite{circuit,susskind}. Describing complexity hollographically has received a lot of interest in the literature thses days and there exist two conjectures, namely the CV (complexity=volume) and CA (complexity=action). In CA conjecture, the complexity is given by the bulk action evaluated on the Wheeler-de Witt patch anchored at some boundary time \cite{Brown:2015lvg,Brown:2015bva}. The CV proposal states that the complexity is identified as the volume of the extremal/maximal volume of a codimensional-one hypersurface $\mathcal{B}$ in the bulk ending on a time slice of the boundary
\begin{equation}
{\cal{C}}=\frac{V(\mathcal{B})}{\hat{L} G_N},
\label{cv}
\end{equation} 
where $G_N$ is the five-dimensional gravitational constant and $\hat{L}$ is some length scale of the bulk, for example the $AdS$ radius. Inspired by the Hubney-Ryu-Takayanagi proposal, this conjecture for the whole boundary system is generalized for subsystems \cite{alishahiha}. The complexity for a subsystem $A$ on the boundary equals to the volume of codimensional-one hypersurface enclosed by Hubney-Ryu-Takayanagi surface, $\gamma_A$, 
\begin{equation}
{\cal{C}}_A=\frac{V_{\gamma_A}}{8\pi R G_N},
\label{sc}
\end{equation}
where $R$ is $AdS_5$ radius and ${\cal{C}}_A$ is known as holographic subregion complexity (HSC). Some recent works on CA and CV prescription and HSC for different gravity models can be found in  \cite{volume,comments,Zhang:2017nth,Zhang:2019nth,quench,renormalization,faregh,mozaffar,mozaffar2,mansoori,mozaffar3, Ebrahim:2018uky}.
In this paper we would like to compute HSC to study the confinement-deconfinement phase transition and the questions we are indeed interested in are: is HSC a relevant order parameter of the phase transition or can HSC recognize the favorable vacuum? or more generally what do we learn about the phase transition using HSC as a quantity we can calculate. To do so, we consider a background which almost perfectly describes the potential in the confined phase  \cite{Andreev:2006ct}. Moreover, using entanglement entropy, the critical temperature $T_c$ can be excellently found in this background \cite{ahmad}. Thus, we start introducing the background(s) and then compute the HSC on different relevant background(s) and discuss its properties to answer above questions.\\

{\textit{\textbf{Backgrounds}}}:
In \cite{Andreev:2006ct}, a five dimensional metric, called modified $AdS\ (MAdS)$, is defined as 
\begin{equation}
ds^2=\frac{R^2}{z^2}g(z)\left(-dt^2+d\vec{x}^2+dz^2\right),
\label{metric}
\end{equation}
and the black hole version of above metric, which we call modified black hole $(MBH)$, is introduced by
\begin{equation}
ds^2=\frac{R^2}{z^2}g(z)\left(-f(z)dt^2+d\vec{x}^2+\frac{dz^2}{f(z)}\right),
\label{metric2}
\end{equation}
where $g(z)=e^{\frac{c}{2}z^2}$, $\vec{x}\equiv (x_1, x_2, x_3)$ and $z$ is the radial coordinate. $c$ is a modifier parameter with energy$^2$ dimension. From the fit to the slope of the Regge trajectories, it is estimated to be $0.9$ GeV$^2$ \cite{Andreev:2006ct}. This number can be found more precisely by determining the critical temperature and it turns out to be $0.94$ GeV$^2$\cite{ahmad}. Clearly the backgrounds \eqref{metric} and \eqref{metric2} are asymptotically $AdS_5$ with radius $R$ and the holographic QCD model is living on the boundary of the backgrounds located at $z=0$. 

The background \eqref{metric2} is a natural extension of \eqref{metric} for including thermodynamics where $f(z)=1-z^4/z_h^4$. The Hawking Temperature is given by $T=\frac{1}{\pi z_h}$ and $z_h$ is the position of the horizon. By setting modifier parameter $c$ equals zero in backgrounds \eqref{metric} and \eqref{metric2}, one can easily find $AdS_5$ background $(AdS)$ and $AdS_5$ planar black hole metric $(BH)$, respectively. The phase transition, in gravitational picture, is described by changing the background geometry from \eqref{metric} to \eqref{metric2} and we will see that HSC does confirm it near the critical point! More properties of these geometries have been discussed in \cite{Andreev:2006ct,Andreev:2006eh,ahmad,Andreev}. \\

{\textit{\textbf{{Potential Energy}}}:
In quantum field theory, potential energy between a quark and antiquark (meson) can be obtained from the expectation value of the Wilson loop, as a non-local gauge invariant operator. In fact, it can be done by evaluating the expectation value of Wilson loop on a rectangular loop, ${\cal{R}}$ with two sides, time $\cal{T}$ and distance $r$, that ${\cal{T}}\gg r$ (see figure \ref{fig00} by the replacement $L\rightarrow {\cal{T}}$ and $l\rightarrow r$). This configuration is equivalent to a static pair of quark-antiquark with the distance $r$ between them.
It is well-known that the expectation value of Wilson loop is dual to the on-shell classical action $S({\cal{R}})$ of a classical string whose worldsheet ends on the $\cal{R}$, rectangular loop at the boundary \cite{Maldacena:1998im}.
It is then straightforward to find potential energy between the pair in the background \eqref{metric} and it turns out \cite{Andreev:2006ct}
\bea\label{maximum} %
V(r)=\left\{%
\begin{array}{ll} %
p\left(-\frac{\kappa_0}{r}+\sigma_0 r+ O(r^3)\right), \ \ \ \ \ r\rightarrow 0 \\
p(\sigma r), \ \ \ \ \ \ \ \ \ \ \ \ \ \ \ \ \ \ \ \ \ \ \ \ \ \ \ \ r\rightarrow\infty\\
\end{array}%
\right.
\eea %
where $p\approx0.94$, $\kappa_0\approx0.23$, $\sigma_0\approx0.16$ GeV$^2$ and $\sigma\approx0.19$ GeV$^2$ for $c=0.9$ GeV$^2$. This potential is smilar to Cornell potential and gives the expected linear and $1/r$ behavior at large and short distance, respectively. In other words, this background describes the QCD at low energy, confined phase. For more details see \cite{Andreev:2006ct}.\\

{\textit{\textbf{Entanglement Entropy}}}: A well-known non-local observable in the information theory is entanglement entropy \cite{thesis}. Consider a quantum field theory whose pure state is described by the density matrix $\rho$. The entanglement entropy of a spatial subregion $A$, with complement $\bar{A}$, denotes how much entanglement exists between $A$ and $\bar{A}$ and it is given by
\be %
S_A=-Tr(\rho_A \log\rho_A),
\ee %
where $\rho_A=Tr_{\bar{A}}(\rho)$ is reduced density matrix and obtained by tracing over the degrees of freedom in the region $\bar{A}$. The entanglement entropy of $A$ shows the amount of information lost when an observer is limited to the subregion $A$. Although calculating the entanglement entropy is normally difficult, the holographic method provides a handy approach to obtain the entanglement entropy. In fact,
Ryo and Takayanagi first proposed in \cite{Nishioka:2009un} that the entanglement entropy can be computed from
\be \label{rt}%
S_A=\frac{\rm{Area}(\gamma_A)}{4G_N},
\ee %
where $\gamma_A$ is a codimension-2 minimal surface whose boundary $\partial\gamma_A$ coincides with the boundary of the subregion $A$ on the boundary of the bulk where the quantum field theory lives, i.e. $\partial\gamma_A=\partial A$. This proposal received a lot of interest during the last decade and passed several non-trivial checks known in the quantum field theory. For more details, we refer the interested reader to \cite{Takayanagi}.
\begin{figure}[H]
\centering
\includegraphics[width=80mm]{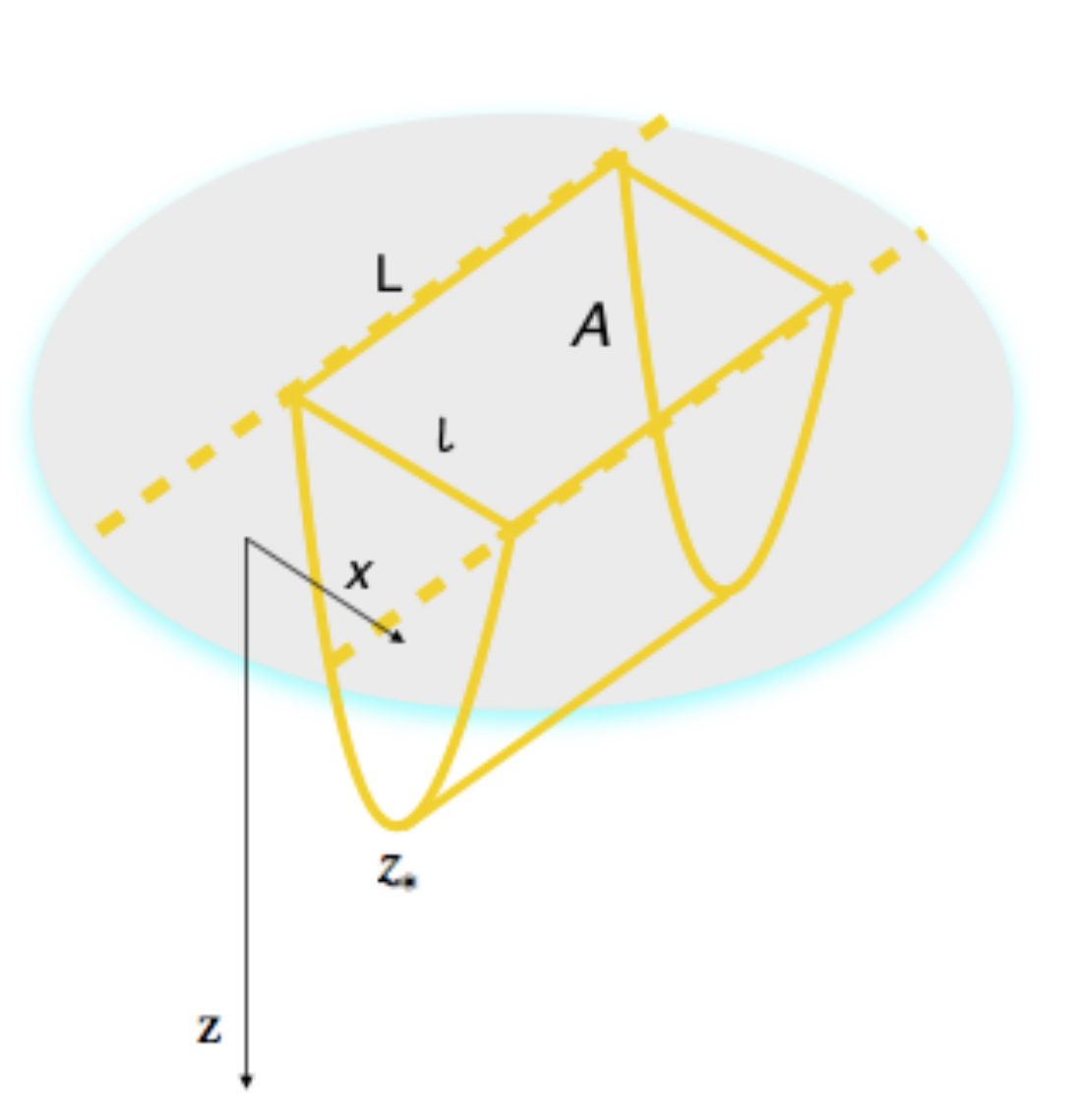} 
\caption{A strip entangling surface of length $l$ and width $L\rightarrow\infty$.}
\label{fig00}
\end{figure} 
Now let us consider a strip entangling surface of length $l$ and width $L\rightarrow\infty$. Indeed the subsystem $A$ is defined by $-\frac{l}{2}<x_1(\equiv x)<\frac{l}{2}$ and $x_2, x_3\in(-\infty,+\infty)$ at a given time, see figure \ref{fig00}. Entanglement entropy is proportional to the minimal area of $\gamma_A$ and it turns out to be \cite{ahmad}
\begin{equation}
S_A^{(c)}=\frac{L^2}{2G_5}\int_{0}^{z_*}\frac{R^3z^3g(z)^3}{z_*^3\sqrt{z_*^6g(z)^3-z^6g(z_*)^3}}dz,
\label{entanglement}
\end{equation}
where $z_*=z(x=0)$. Considering disconnected configuration described by two surfaces located at $x=\pm l/2$, the entanglement entropy for disconnected case becomes \cite{ahmad}
\begin{equation}
S_A^{(d)}=\frac{L^2}{2G_5}\int_{0}^{\sqrt{\frac{2}{c}}}{\left(\frac{R^2g(z)}{z^2}\right)}^{\frac{3}{2}}dz.
\label{disconnected}
\end{equation}
Then in order to specify the phase transition, according to the results reported in \cite{ahmad}, we define
\be\label{EE}
\Delta S(l)\equiv\frac{2G_5}{L^2}\left(S_A^{(c)}-S_A^{(d)}\right).
\ee 
Using the above definition the phase transition and its critical temperature has been predicted \cite{ahmad}. It is shown that the phase transition occurs at $l_c\approx1$ fm (compatible with the size of the hadrons) for the reasonable value of c, i.e. $c\approx 0.9$ GeV$^2$.
As shown in figure \ref{fig2}, blue curve, the point at which the sign of $\Delta S$ changes characterizes the phase transition. It occurs in the regime of $l$ for which the linear behavior of the Cornell type potential has been dominated.\\



{\textit{\textbf{Subregion Holographic Complexity}}}:
Motivated by holographic entanglement entropy and by an extension of the Hubney-Ryu-Takayanagi proposal, HSC for a subsystem $A$ in the boundary theory is defined as follows \cite{alishahiha} 
\begin{equation}
{\cal{C}}=\frac{V_{\gamma_A}}{8\pi R G_5},
\label{co}
\end{equation} 
where $V_{\gamma_A}$ is the volume of the codimension-one hypersurface enclosed by minimal hypersurface $\gamma_A$ obtained to calculate holographic entanglement entropy.
Then using metric \eqref{metric}, one easily finds the the area of the minimal surface 
\be %
S=\frac{L^2}{4G_5}\int_{-l/2}^{l/2}\frac{R^3}{z^3}g(z)^{\frac{3}{2}} \sqrt{1+z'(x)^2}dx,
\ee %
where $z(x)$ (or equivalently $x(z)$) is the profile of the minimal surface. Then by using the constant of motion, the
profile of the minimal surface is obtained
\begin{equation}
x(z)=2\int_z^{z_*}\frac{z^3g(z_*)^\frac{3}{2}}{\sqrt{z_*^6g(z)^3-z^6g(z_*)^3}}dz.
\label{profile}
\end{equation}
 In the static case the volume enclosed by $\gamma_A$ is obtained by integrating the inside of the minimal surface. It can be done by slicing the bulk with planes of constant $z$. We therefore have
\begin{equation}
V_{\gamma_A}(z_*)=2L^2\int_0^{z_*}\frac{R^4}{z^4}g(z)^2x(z)dz,
\label{volume}
\end{equation}
and similar calculation for the metric \eqref{metric2} leads to 
\begin{equation}
V_{\gamma_A}(z_*)=2L^2\int_0^{z_*}\frac{R^4}{z^4}\frac{g(z)^2x(z)}{\sqrt{f(z)}}dz.
\label{volume}
\end{equation} 
It is obvious that the above two equations reduce to the case of $AdS$ and $BH$ by setting $c=0$, respectively.
Similar to the case of entanglement entropy, the volume is divergent and we need to introduce a normalized volume, {\it{relative complexity}}, as follows 
\begin{subequations}
\begin{align}
& \label{eq1} C_1\equiv\frac{{\cal{C}}_{MAdS}}{{\cal{C}}_{AdS}}-1,\\
& \label{eq3} C_2\equiv\frac{{\cal{C}}_{MBH}}{{\cal{C}}_{BH}}-1,
\end{align}
\end{subequations}
where ${\cal{C}}_{MAdS}$, ${\cal{C}}_{AdS}$, $ {\cal{C}}_{MBH}$ and ${\cal{C}}_{BH}$ are the HSC for $A$ in $MAdS$, $AdS$, $MBH$ and $BH$ geometry, respectively. We use the above definitions to discuss the phase transition. \\

{\textit{\textbf{Numerical results}}:
In this section we will argue that our findings from the numerical calculations of HSC at zero as well as non-zero temperature are reasonable. Indeed, a general expectation is that the HSC should behave differently at zero and non-zero temperatures since each of them indicates the confinement and non-confinement regime in field theory, respectively.\\

{\textit{\textbf{Zero temperature}}:
As it was previously mentioned, the state we consider here corresponds to the modified background \eqref{metric} probed by a classical string (quark-antiquark pair or meson in the dual field theory) on the gravity side. Therefore, according to the definition \eqref{co}, ${\cal{C}}_{MAdS} ({\cal{C}}_{AdS})$ is identified with the complexity of the probe meson living in non-conformal (conformal) vacuum both at zero temperature. Our results are shown in figure \ref{fig2}, left. The red and blue curves denote the potential energy between the pair and difference of entanglement entropies $\Delta S$, respectively. The black curve is the relative complexity of the pair obtained from \eqref{eq1}. The horizontal axis is the distance between quark-antiquark pair, i.e. $r\equiv l$. As it is clearly seen, the relative complexity increases with $l$ up to a maximum at $l=l_{max}$ and then decreases with rising $l$. It is important to notice that it changes sign, let's say, at $l=l_s$. There are obviously three different regions as follows:
\begin{itemize}
\item $C_1>0\ {\rm{and}}\ l<l_s$: meaning that ${\cal{C}}_{MAdS}>{\cal{C}}_{AdS}$ for small values of $l$, $i.e.\ l<l_s$, probing the UV regime in field theory. To be more specific, at high energy the information needed to prepare the state of meson in the non-conformal vacuum is larger than in the conformal one.
\item $C_1<0\ {\rm{and}}\ l_s<l<l_c$: It means that ${\cal{C}}_{MAdS}<{\cal{C}}_{AdS}$ and therefore the state of meson is easier to specify in the non-conformal vacuum than the other one.
\item $C_1<0\ {\rm{and}}\ l>l_c$: It is similar to the previous case.
\end{itemize}

\begin{figure}[H]
\centering
\includegraphics[width=80 mm]{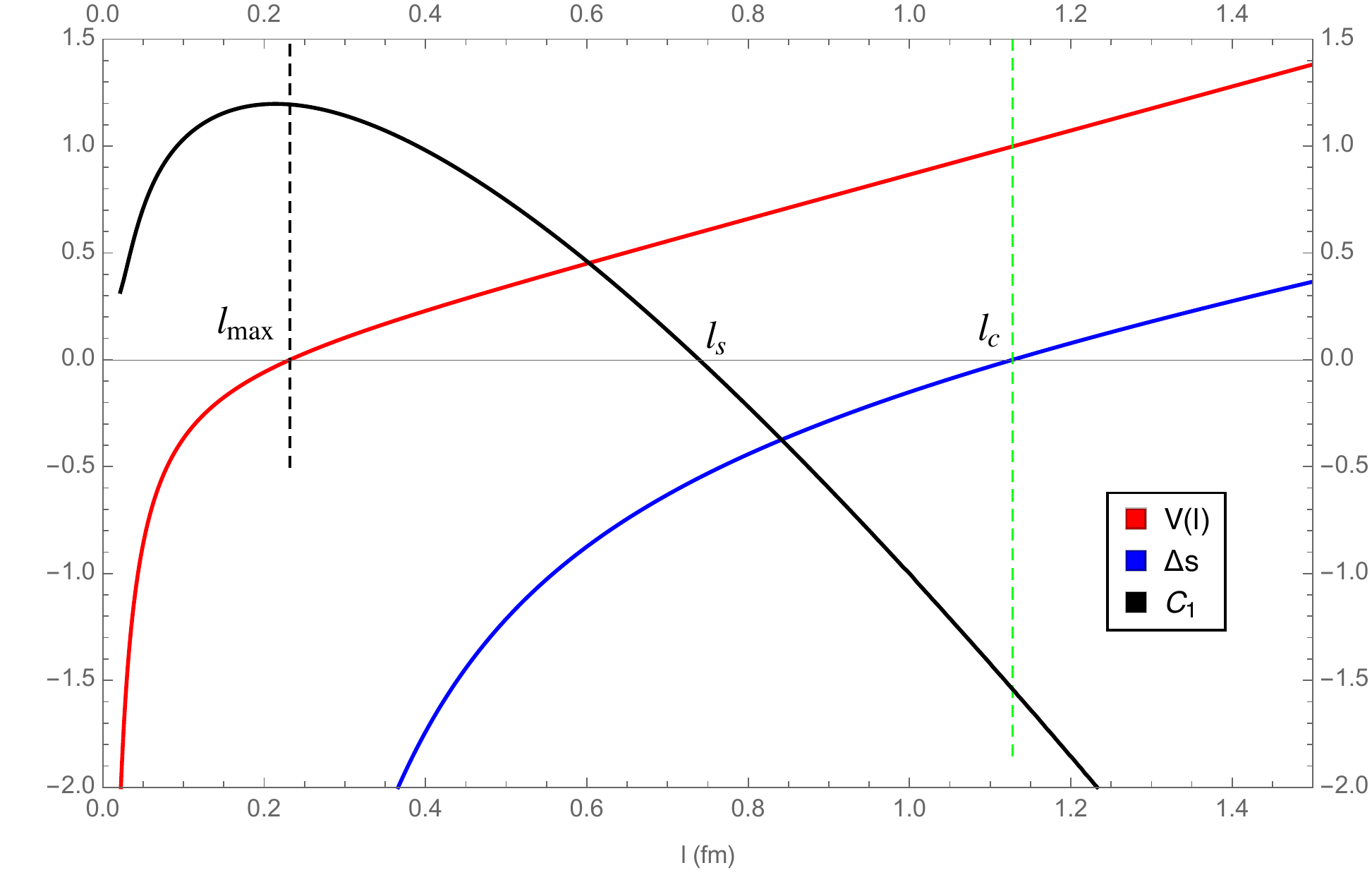} 
\includegraphics[width=80 mm]{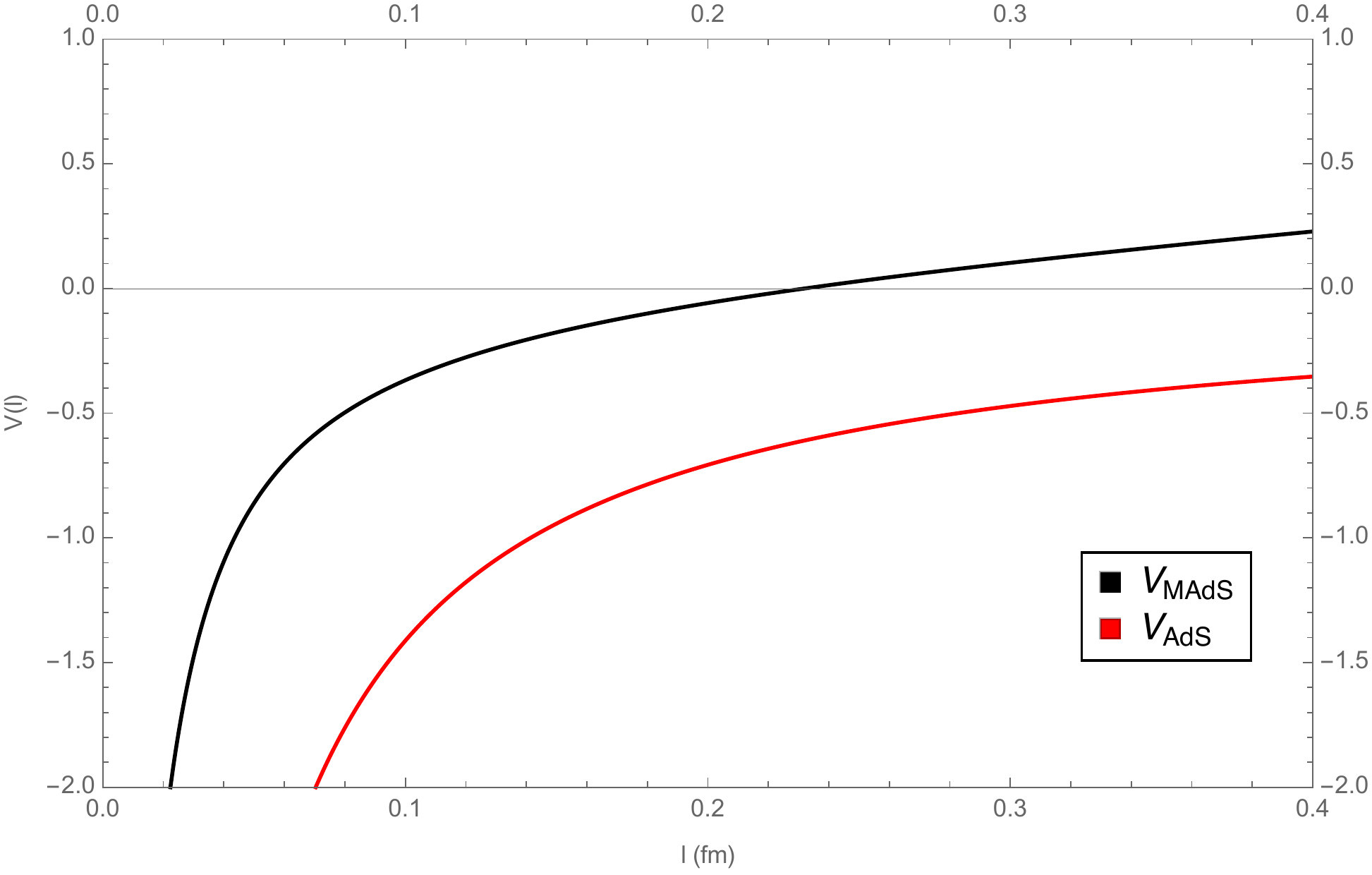} 
\caption{Left: The potential energy $V(l)$, the difference of the entanglement entropies $\Delta{S}$ and $C_1$ in terms of $l$ (fm) for $c=0.94$ GeV$^2$. Right: The potential of meson in $AdS$ and $MAdS$ background for $c=0.94$ GeV$^2$.}
\label{fig2}
\end{figure} 

These above categories have a simple physical interpretation as follows. Based on results reported in \cite{ahmad}, the QCD phase transition happens in the region with $C_1<0$ and exactly at $l=l_c$ and $T_c=l_c^{-1}=175$ MeV. Therefore, the region with $l<l_c$ describes the field theory in the confined phase and the region of $l>l_c$ indeed is not physical and the background \eqref{metric} must be replaced with the black hole one which, in this case, is \eqref{metric2} or $AdS_5$ planar black hole. As a result, the third category is not acceptable. Moreover, for allowed values of $l$ and near $l_c$ in the confined phase, second category, the complexity of the meson in the non-conformal vacuum is less than the vacuum one or equivalently we need less information to prepare the meson in the non-conformal vacuum. It seems reasonable since the background \eqref{metric} describes confined phase of QCD and produces the relevant potential energy between the pair given by \eqref{maximum} in agreement with our intuition. Put in other words, although HSC does not help us to find the value of critical temperature, it states that for $E(=l^{-1})\lesssim T_c$ the state of the meson in the non-conformal vacuum is easier to specify.

About the first category, we would like to make a comment on the potential energy. It can be seen from this figure that the potential energy of meson becomes zero around $l_{max}$ at which the maximum value of the relative complexity $C_1$ happens. Further, it is significant to notice that the difference between conformal and non-conformal vacuum is expected to realize more by larger values of $l$ and as a result, the relative complexity should be a monotonically increasing function. However, when the $r$ term in the potential starts dominating, the relative complexity $C_1$ decreases! Therefore one may conclude that at zero temperature, as soon as the $r\ (\frac{1}{r})$ term is dominant in the potential energy, the relative complexity decreases (increases). In fact, at least in this model, increase or decrease of relative complexity $C_1$ seems to be related to the dominant term in the potential energy.


Our results become more interesting when we consider the stability of meson at zero temperature in conformal and non-conformal vacuum too. The potential energies for both cases have been plotted in figure \ref{fig2}, right. As this figure shows, a meson in the non-conformal vacuum is always less stable than conformal one. Moreover, our complexity calculation indicates that the less stable meson needs less information to spacified, of course near the critical point, $175$ MeV $\lesssim E \lesssim 270$ MeV. It is therefore seems that near the critical point a better option is to choose the less stable meson (in a non-conformal vacuum) with less information. However, for $l<l_s$ the story is vice versa and the less stable meson, needs more information to specified, see table \ref{list}. One may thus conclude that the regime of $E>270$ MeV is much better described by a conformal vacuum (instead of a non-conformal one) since it is more stable and needs less information to specify a meson!

We would like to mention that we intuitively expect that more (less) information is needed to specify a more (less) stable meson. 
Altogether the only conflict arises when $E > 270$ MeV, in the region where $\frac{1}{r}$ term dominates or $r$ term starts dominating in the potential. Then this incompatibility is cured in the regime that $r$ term in the potential plays a more importane role, i.e. close to the critical point! As a matter of fact, it seems that the $r$ term in the potential help us to find the appropriate result corresponding to our intuition. 

\begin{table}[ht]
\caption{A short summary of information and stability in non-conformal vacuum\\
 comparing to conformal one}
\vspace{1 mm}
\centering
\begin{tabular}{c c c c}
\hline\hline
~~$ \rm{Cases} $ ~~   &~~$ \rm{Information} $ ~~   &   ~~ $ \rm{Stability} $ ~~  \\[0.5ex]
\hline
$T=0,\ l<l_s$ & \rm{more} &  \rm{less}  \\
$T=0,\ l_s<l<l_c$ & \rm{less} & \rm{less}   \\
$T\ne 0,\ l>l_c$ & \rm{more} &  \rm{more}  \\
\hline
\end{tabular}\\[1ex]
\label{list}
\end{table}

Since complexity refers to classifying various quantum states based on their difficulties, it can be also defined as a difficulty in creating a state \cite{susskind}. From this point of view, the above results indicate that, near the critical temperature, the difficulty in preparing a meson state in the non-conformal vacuum, dual to \eqref{metric}, is less than the conformal. 

\begin{figure}[H]
\centering
\includegraphics[width=80 mm]{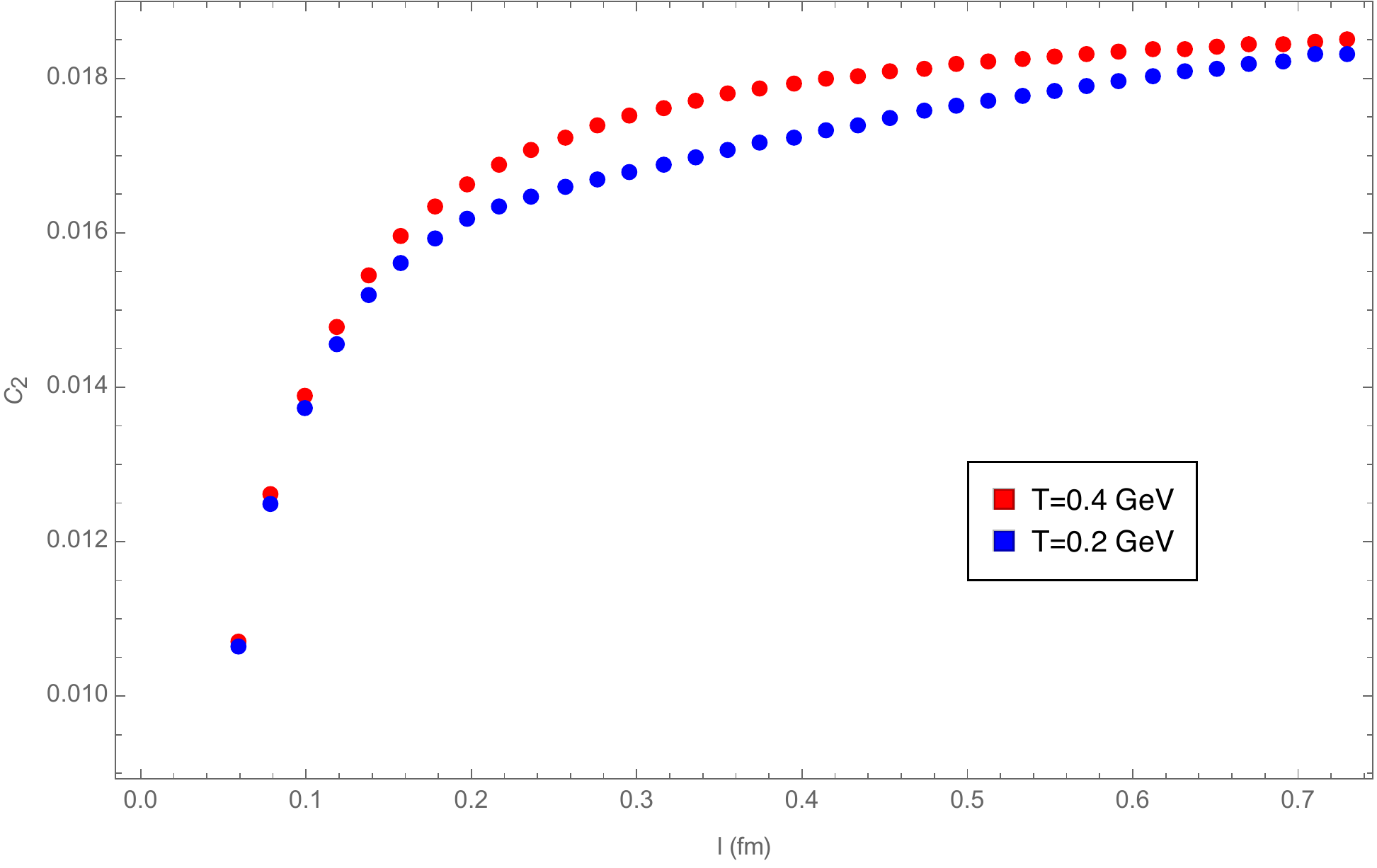} 
\includegraphics[width=80 mm]{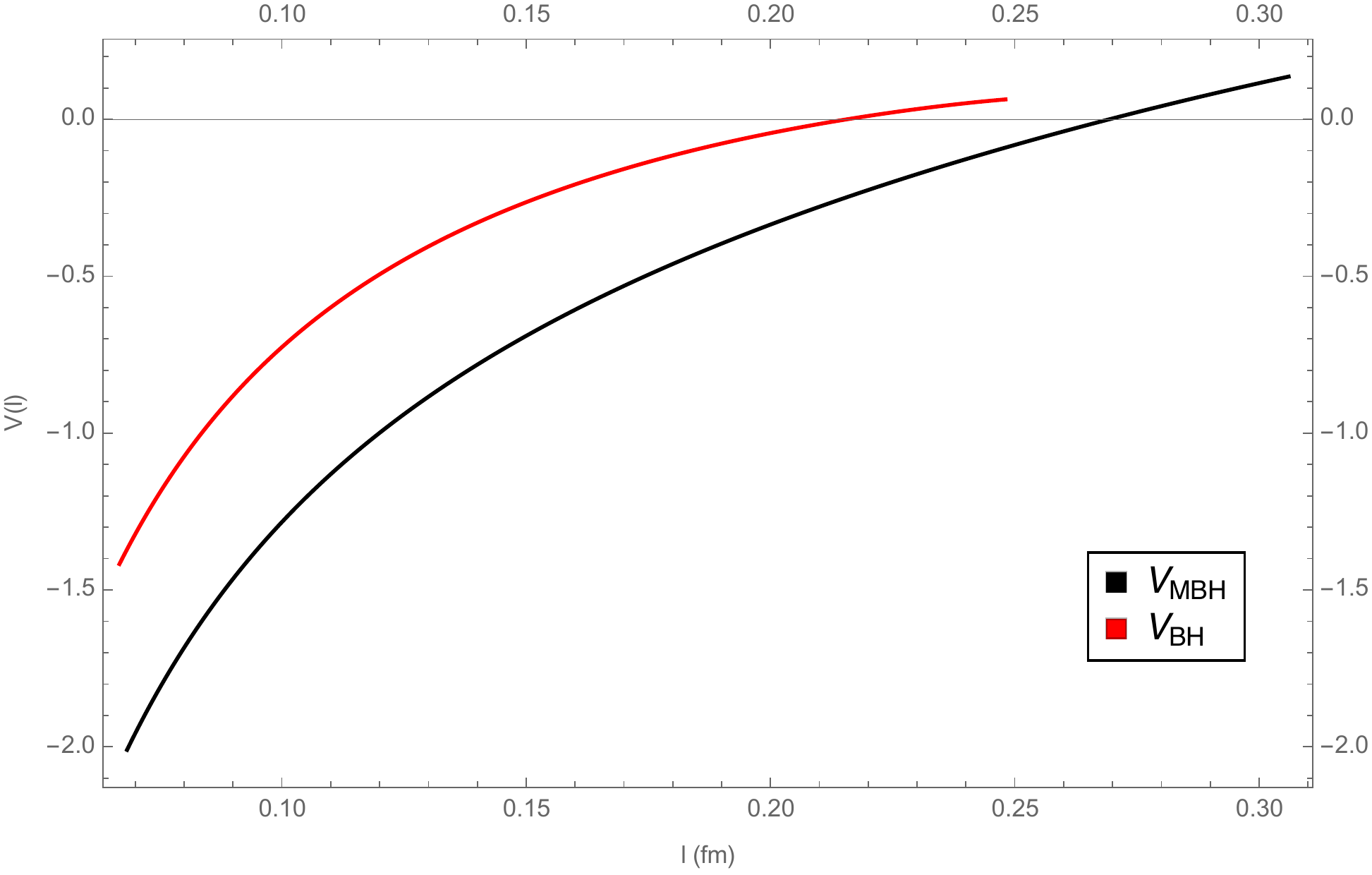} 
\caption{Left: $C_2$ as a function of $l$ (fm) for $c=0.94$ GeV$^2$. Right: The potential energy of meson for $BH$ and $MBH$ for $T=$ 200 Mev.}
\label{fig3}
\end{figure} 

{\textit{\textbf{Finite Temperature}}}:
Similar to the zero temperature case, we plot $C_2$ in terms of $l$ in figure \ref{fig3}, left. We observe that there is no substantial difference among various temperatures and also $C_2$ is always positive, i.e. ${\cal{C}}_{MBH}>{\cal{C}}_{BH}$. Therefore, in order to specify the state of meson in the thermal vacuum, less information is needed than in the non-conformal thermal vacuum. It is easy to check that for a given value of $l$ quark and antiquark pair is less bounded in the thermal vacuum, as one can see in figure \ref{fig3}, right and thus the bound state in the thermal non-conformal vacuum is stronger. In other words, the color screening, which prevent quark and antiquark from binding to each other in the deconfined phase, is stronger in the thermal vacuum. Therefore, in order to specify a bound state in the thermal non-conformal vacuum, we need more information since the quark and antiquark are more bounded.

\begin{figure}[H]
\centering
\includegraphics[width=88 mm]{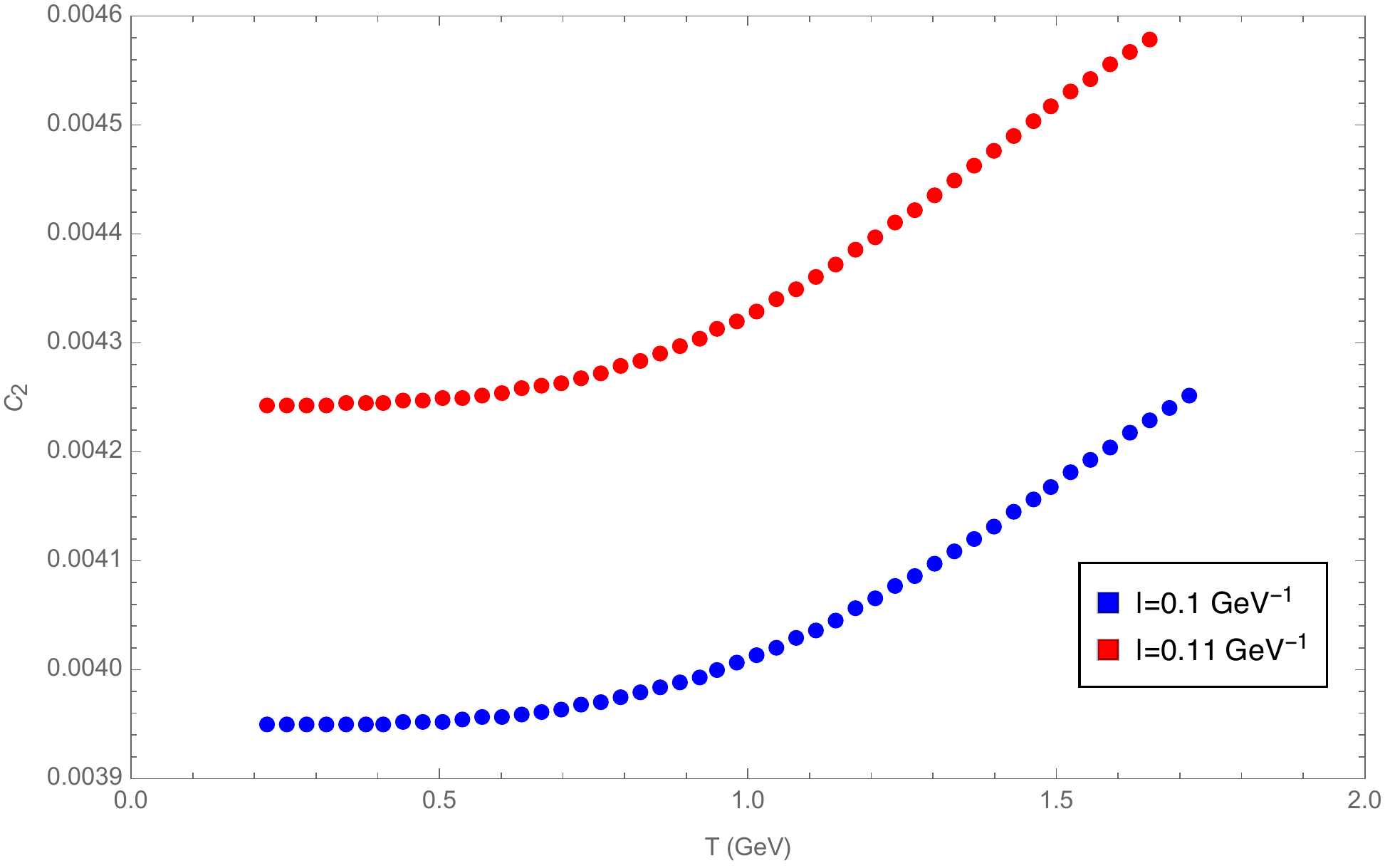} 
\caption{$C_2$ as a function of $T$ for $c=0.94$ GeV$^2$.}
\label{fig4}
\end{figure} 

We would also like to mention that, based on figure \ref{fig2} and its discussion, the potential energies in figure \ref{fig3}, right are obviously not linear and we naively expect the relative complexity $C_2$ increases similar to figure \ref{fig2} and it is confirmed by figure \ref{fig3}, left. Therefore, it seems that in a given vacuum a connection between the complexity of meson and the potential energy of the pair exist. In fact, it can be classified in two categories as follows:
\begin{itemize}
\item When the linear term is not as important as the $r$ term in the potential energy the absolute value of relative complexity increases. This result is confirmed by the relative complexity $C_1$ and $C_2$. Note that although, for instance, the relative complexity $C_2$ is not negative, this positivity has no physical interpretation. Because one can define  $C'_2\equiv\frac{{\cal{C}}_{BH}}{{\cal{C}}_{MBH}}-1$ which is clearly negative and leads to the same results. Moreover, it seems that the place of the maximum (or minimum) depends on the strength of the potential energies though we do not have any analytical or numerical calculation.
\item Otherwise, when the $r$ term is dominant the absolute value of relative complexity decreases.
\end{itemize}

In figure \ref{fig4} we plot $C_2$ in terms of temperature for two values of $l$ to cross-check our previous result about the color screening. It is obvious that for larger value of $l$ we need more information to specify the state of probe meson in a non-conformal thermal vacuum. In other words, the screening becomes stronger when the temperature or distance $l$ raises. Therefore, as we discussed already, the screening of the static quark-antiquark potential and $C_2$ together increase.\\

\textit{\textbf{Acknowledgment}}:
M. A. would like to thank CERN TH-Division for 
warm hospitality.

\end{document}